\documentclass[a4paper]{PoS}

\title{Centrality Issues In Asymmetric Collisions: Direct Photons To The
Rescue?}

\ShortTitle{Centrality Issues In Asymmetric Collisions: Direct Photons To
The Rescue?}

\author{\speaker{Gabor David}%
        \thanks{A footnote may follow.}\\
       Stony Brook University\\
       E-mail: \email{david@bnl.gov}}

\abstract{Methods to classify events experimentally according to
  collision geometry are well established and non-controversial when
  collisions of large ions are studied.  However, high luminosity data
  from $p/d$+A collisions at RHIC and LHC provided some surprising
  results that either call for new physics or question the
  applicability of the established methods of event classification for
  them. 
  So far there is no consensus in the community what is the proper model
  and procedure to determine centrality, how to connect observed event
  activity with collision geometry in very asymmetric, 
  $p/d$+A collisions in the same sense and with the same accuracy as          
  was done in A+A.  We argue that high $p_T$ direct photons offer an
  {\it a posteriori} test of any method suggested to categorize $p/d$+A
  -- and in general, very asymmetric -- collisions: the method is only
  viable if the nuclear modification factor for high $p_T$ direct
  photons is about unity for all centrality classes.
}

\FullConference{The 26th International Nuclear Physics Conference\\
                 11-16 September, 2016\\
                 Adelaide, Australia}

\begin{document}

\section{Introduction}

Establishing collision geometry from experimental observables when the 
internal substructure of the colliding objects is relevant, is an old
problem in particle and nuclear physics.  Here ``collision geometry''
can be something as simple as the impact parameter $b$, or the shape
of the overlap region of the two objects (particles or nuclei), the
spatial distribution of the components (partons or nucleons) therein,
the number of components in the two objects that participate in some
interaction (parton or nucleon participants, $N_{part}$), finally the
total number of such interactions (``binary collisions'', $N_{coll}$).
Since none of these are directly observable, the transition
from $b$ to $N_{part},N_{coll}$ requires some theoretical model.
The issue was first addressed on the nucleon (hadron) level by 
Glauber~\cite{glauber1959}.  His model assumes incoherent collisions
of the nucleons moving on a straight path, constant $\sigma_{NN}$ and
small momentum transfer in the individual $NN$ collisions
({\it ``...the approximate wave function (74) is only adequate for the
  treatment of small-angle scattering.  It does not contain, in
  general, a correct estimate of the Fourier-amplitudes corresponding
  to large momentum transfer''}). Despite these limitations the
Glauber-model is successfully used in relativistic heavy ion
collisions to establish $N_{part}$, $N_{coll}$ and to connect the 
impact parameter $b$ (or ``centrality'') to experimentally measurable
quantities -- at least when the colliding ions are both large.  The
procedures are described for instance in~\cite{reygers2007} including
the caveat that {\it ``in heavy ion collisions we manipulate the fact
  that the majority of the initial state nucleon-nucleon collisions
  will be analogous to minimum bias $p$+$p$ collisions''}.  Recently
is has been observed~\cite{ppg100} that bulk observables, like
$dE_T/d\eta$ are better described if the fundamental interactions happen
between constituent-quarks rather then nucleons, but this approach
still preserves the basic characteristics of the Glauber model
(incoherent scatterings, straight path, constant $\sigma_{qq}$).
If $N_{qp}$ is the number of constituent quark participants, the mean
transverse energy $<dE_T/d\eta>/N_{qp}$ is approximately constant for
a wide range of collision energies, at least when the colliding ions
are both large.  Even if there are occasional large fluctuations in
an individual nucleon-nucleon (or parton-parton) scattering, their effect
is washed out by the large number of average collisions between
constituents in the event.

The idea of fluctuating cross-sections has been introduced by
Gribov~\cite{gribov1970} and gained traction as collision energies
increased and also with the study of asymmetric systems - $p$+A at
first, followed by light-on-heavy ion collisions.  One of the first
examples was the series of transverse energy measurements by the NA34
(HELIOS) collaboration using $^{32}$S beams on various targets, from
Al to U~\cite{helios1989,helios1991}.  With increasing target size the
tails of the $E_T$ distributions exceeded more and more the
expectations from independent nucleon-nucleon collisions; the excess
was attributed to fluctuations in $E_T$ production, characterized by an
empirical parameter $\omega$ that increases monotonically with target size.
This parameter has been tied to cross-section fluctuations
in~\cite{baym1991}, due to the (frozen) initial configuration of the
nucleons.  This opened the way to a modification of the original
Glauber-model, including the calculation of $N_{coll}$.  However,
experiments at RHIC and LHC continued using the original Glauber-model
to determine collision centrality, $N_{part}$ and $N_{coll}$ until the
early 2010s.  It is interesting to note that in the 2007 review paper on
``Glauber Modeling in High Energy Nuclear
Collisions''~\cite{reygers2007} cross-section fluctuations are not
discussed yet, not even as a footnote.  This situation quickly changed
once experiments at RHIC and LHC started to take large amounts of
p/d+A data, originally meant to fine-tune our understanding of the
initial state, impact-parameter dependent nuclear PDFs, and cold
nuclear matter effects in general.  Instead, some very unexpected
results were found, primarily for the centrality dependence of nuclear
modification factors, which left only two (not mutually exclusive)
possibilities.  Either some new physics processes, so far not seen,
have to be considered - or we have to re-think how collision geometry
can be determined from experimental observables.

\section{Centrality, nuclear modification factor -- A+A collisions}

In heavy ion experiments the collision geometry is usually implied
from some global observable, like charged particle production ($N_{ch}$),
total transverse energy ($E_T$), sometimes also by counting the number 
of nucleons (often $n$ only) that didn't take part in any interaction
(spectators).  Both $N_{ch}$ and $E_T$ are dominated by particles
coming from soft interactions.  The measurement is usually (but not
always) done far away in (pseudo)rapidity from the region where the
centrality-dependent signals will be studied, in order to minimize
auto-correlations.  The total distribution is then divided up to
percentiles, providing the ``centrality'' classification of the
event.  The connection to the directly inaccessible
$b$, $N_{part}$, $N_{coll}$ is then made with a Glauber-model based
Monte Carlo~\cite{reygers2007}.  For each participating nucleon (quark)
the contribution is modeled with a negative binomial distribution
(NBD), tuned such that its convolution with the calculated $N_{part}$
distribution reproduces the measured $N_{ch}$ or $E_T$.  Simulations
show that the method works well when large ions collide: the
correlation between $N_{ch}$ and $N_{part}$ (and $b$ or $N_{coll}$) is 
tight.  This is also confirmed by the nuclear modification factors.

In general terms the nuclear modification factor for an observable 
$X$ (particle species, jet) and nuclei $A,B$ is defined as
$$ R^X_{BA} = \frac{dN^X_{BA}/dp_Tdy}{<N_{coll}>dN^X_{pp}/dp_Tdy}$$
\noindent
{\it i.e.} the ratio of the yield observed in the heavy ion collision
and the yield in $p$+$p$ scaled by the average number of (binary)
nucleon-nucleon collisions.  If  $R^X_{BA}\approx 1$ it is usually
interpreted as the absence of any specific nuclear (or medium)
effects -- although this is clearly a necessary condition only, not a
sufficient one.  If a strongly interacting medium is formed in the
collision, the partons are expected to lose energy therein, making 
$R^X_{BA}< 1$ at higher $p_T$, where hard scattering (i.e. early)
processes are expected to be the dominant production mechanism.  
Such suppression has indeed been observed for various hadrons and 
jets at all RHIC and LHC experiments.  Of course the absolute value of 
$R^X_{BA}$ depends on $N_{coll}$, calculated from a Glauber-model,
which in turn is connected with the experimental centrality by soft
particle production, in a different rapidity region, also, depending 
on $N_{part}$ rather than $N_{coll}$.

\begin{figure}
\centering{
\includegraphics[width=0.5\linewidth]{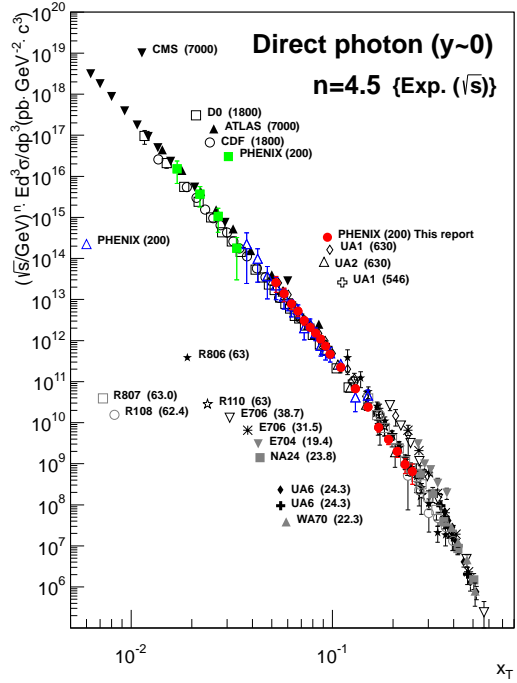}}
\caption{\label{fig:xtscaling}
  Various direct photon cross section measurements in $p$+$p$ and
  $p$+$\bar{p}$ collisions scaled by $(\sqrt{s})^{4.5}$ vs
  $x_T=2p_T/\sqrt{s}$.  The legend shows the experiment and the
  center-of-mass energy (GeV) in parenthesis.  References can be found
  in~\cite{ppg136}. 
}
\end{figure}

Validation of this procedure was ultimately provided by high $p_T$ 
direct photons, predominantly produced in initial hard scattering, but
then, being color-neutral and with $\alpha_{em}<<\alpha_s$ they are passing
through the colored medium virtually unaffected.  This way they are a
good candidate to ``calibrate'' the number of hard collisions, and, by
extension, the $N_{coll}$ calculated from the Glauber model.  Before
showing this, it is worth noting, that their production in
$p$+$p$ is theoretically well understood (see Fig. 6
in~\cite{aurenche2006}).  The $x_T$ scaling of the experimental data
published until 2012 is shown in Fig.~\ref{fig:xtscaling} reproduced
from~\cite{ppg136}.   Over two
orders of magnitude in $x_T$, 13 orders of magnitude in cross-section
and a factor of 350 in $\sqrt{s}$ all data (with the exception of the
controversial Fermilab E706 results) line up on a single
curve, from which an exponent $n=4.5$ can be derived.  Leading order
($2\rightarrow2$) processes only would result in $n=4$; the small
deviation from this value indicates that higher order processes don't
contribute substantially to the photon yield.

As it is well known, in heavy ion collisions all RHIC and LHC
experiments found that in A+A high $p_T$ hadrons are suppressed,
$R^{had}_{AA}$ - calculated with the Glauber $N_{coll}$ - is strongly
dependent on centrality, and usually $R^{had}_{AA}<<1$.  In stark
contrast, for photons $R^{\gamma}_{AA}\approx 1$ was observed
for all centralities (see for instance~\cite{ppg139,cms2012,atlas2016}).
Since $R_{AA}$ has been calculated with the same $N_{coll}$ both for
hadrons (suppression observed) and photons (suppression neither
expected nor observed), this is a potent validation of the Glauber
model in A+A.

It should be noted that strictly speaking $R^{\gamma}_{AA}$ should not
necessarily be unity; there are at least three processes that can
slightly modify it.  The first is photons from jet fragmentation,
where the parent partons already lost energy in the medium; however,
these are only a small fraction of the photons~\cite{ppg136} and often
can be tagged by isolation cuts~\cite{atlas2016}.  The second is
called jet-photon conversion~\cite{fries2003}, when a fast quark
passing through the sQGP produces photons by Compton scattering with
the thermal gluons or annihilation with the thermal quarks.  Photons
from this process, originally thought to be the dominant source up to
5-6\,GeV/$c$, are hard to tag experimentally, but if their rate is
really that high, they should be identifiable from {\it double}
jet-conversion of back-to-back hard scattered partons.  The third
modifying factor is the isospin effect in nucleus-nucleus
collisions~\cite{arleo2006}. 
When calculating $R^{\gamma}_{AA}$ one scales the $p$+$p$
cross-section with $N_{coll}$ of all nucleons, but the photon
production from $p$+$p$, $p$+$n$ and $n$+$n$ is different 
$\Sigma Q^2_q$ quark charge square sum of protons and neutrons.
Despite these three caveats it is safe to say that 1/ production of
high $p_T$ photons in $p$+$p$ is well understood, 2/ $R^{\gamma}_{AA}$
is close to the expected value in large systems (consistent with unity
within experimental uncertainties)  3/ while more precise measurements
may reveal small deviations, those appear to be calculable.  In short,
high $p_T$ photons are ``standard candle'', a tool to calibrate
$N_{coll}$. 

\section{Centrality, nuclear modification factor -- p/d+A collisions}

While hadron and jet suppression in A+A could be explained with the
formation of an sQGP medium (final state effect only), other
observations suggested that the initial (pre-collision) state can also
be modified.  The large data sets collected since 2008 at RHIC
($d$+Au) and later at LHC ($p$+Pb) were originally meant to study
these initial state effects: the expectation was that colliding these
very asymmetric systems (no more than two nucleons on a large ion)
will probe the properties of the ``cold'' nucleus, where no sQGP is
formed, and the results serve as a baseline in the study of medium
effects in A+A.  The centrality of the $p/d$+A collisions was
initially determined by the same methods that worked well for A+A.

The first results were quite surprising.  Observations of long-range
azimuthal correlations and strong azimuthal anisotropies (flow) raise
the possibility that even in these very asymmetric collisions droplets
of sQGP can be formed.  As for the nuclear modification factor
$R_{pA}$ at mid-rapidity, the findings were even more puzzling.  On
the one hand, in ``central'' collisions $R_{pA}$ was suppressed (this in
itself was still consistent with droplets of sQGP), on the other hand
in ``peripheral'' collisions $R_{pA}$ showed significant 
enhancement~\cite{atlas2015,ppg184}, defying all expectations and not seen in 
any previous $R_{AA}$ measurement.  While some new physics mechanism 
producing such enhancement at high $p_T$ could not be excluded, in
Occam's spirit it was only logical to assume that maybe the way
centrality is determined in A+A isn't directly applicable in those
very asymmetric collisions~\cite{david2013}.

\begin{figure}
\centering{
\includegraphics[width=0.8\linewidth]{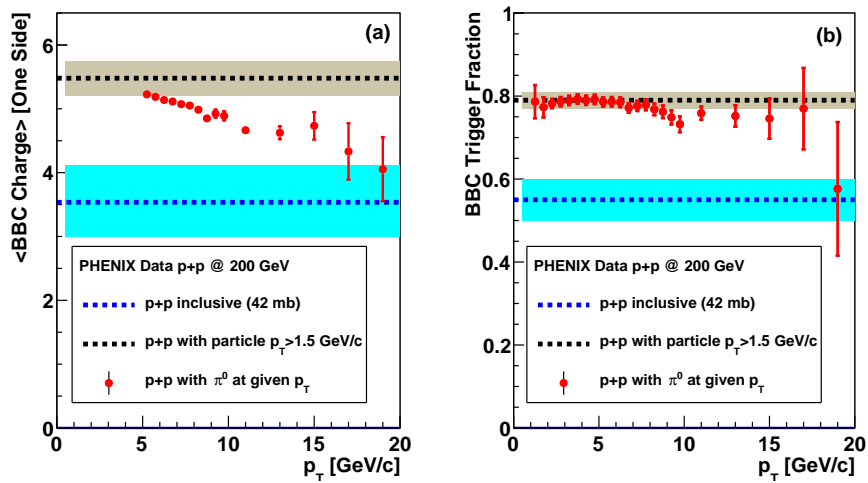}}
\caption{\label{fig:ppg160_ppbiasdata}
  Left (a): $N_{ch}$ at $-3.9<\eta<-3.1$ vs the highest $p_T$ observed
  in a single particle at $|\eta|<0.35$ in $pp$
  collisions~\cite{ppg160}.  The two dashed lines are the mean charge
  for events taken with minimum bias trigger (lower, blue) and requiring
  at least one particle with $p_T>1.5$GeV/$c$ at midrapidity.
  Right (b): Trigger efficiency (probability of the coincidence of at
  least one particle at both $-3.9<\eta<-3.1$ and $3.1<\eta<3.9$) for
  minimum bias events (lower, blue line), events with at least one
  particle with $p_T>1.5$ at midrapidity (upper, black line), and the
  dependence on the highest $p_T$ particle observed at midrapidity.
}
\end{figure}

The strictly empirical argument in~\cite{david2013} was this.  The
only case where the correlation between soft production at high
rapidities (where centrality is usually determined) and high $p_T$
particle/jet production at mid-rapidity can experimentally be verified
without any bias or prior assumption is the case of $p$+$p$
collisions.  In Fig.~\ref{fig:ppg160_ppbiasdata} taken
from~\cite{ppg160} this correlation is presented.  On the left panel
the average $N_{ch}$ in the forward detector (determining centrality
and serving as trigger) is shown {\it vs} the transverse momentum of
the highest $p_T$ particle observed at midrapidity.  The right panel
shows the trigger efficiency {\it vs} the same quantity.  While the
presence of a high $p_T$ particle at midrapidity causes only minor
losses in the trigger efficiency, the depletion in forward $N_{ch}$ is
very substantial and grows rapidly with $p_T$.  This is not an issue
in A+A collisions, since there are many collisions between 
{\it  different} nucleons, and even if one nucleon-nucleon scattering
is hard, contributing less to the $N_{ch}$ at high rapidity, the
deficit is virtually invisible since all other nucleon-nucleon
collisions produce the average $N_{ch}$, and the centrality calculated
with the Glauber model remains unbiased.  The same is not true in
$p/d$+A collisions: once the projectile suffers a hard collision,
$N_{ch}$ is necessarily depleted.  Even if its unaffected constituents 
have further interactions with nucleons in the target A, the total
$N_{ch}$ will shift to lower values.  As a consequence, an
event with a high $p_T$ particle will be classified on the average as 
more peripheral than it actually is.  This simple, qualitative picture
has the advantage that it relies only on actual experimental
observations, and it is consistent both with the apparent suppression
of $R_{pA}$ in ``central'' and its apparent enhancement in ``peripheral''
$p/d+A$ collisions.  It doesn't claim to provide an explanation of the
underlying physics mechanisms.

In the past few years there were many attempts to modify the Glauber
procedure for very asymmetric collisions based on some
phenomenological model.  In an early paper the Gribov-picture and
the notion of color fluctuations has been
re-introduced~\cite{alvioli2013}.  The authors found that in $p$+A
collisions ``standard procedures for selecting peripheral (central)
collisions lead to selection of configurations in the projectile which
interact with smaller (larger) than average strength''.  
The authors of~\cite{alvioli2014} explicitely studied $N_{coll}$ in 
case of hard
triggers and assuming the ``flickering'' of the interaction strength 
in $p$+A collisions, finding that ``measurements by CMS and ATLAS for
jets carrying a large fraction of the proton momentum, $x_p$, is
consistent with the expectation that these configurations interact
with the strength that is significantly smaller than the average
one''.  The authors of~\cite{bathe2016} provide a model in which the
removal of a large $x$ parton (the one producing the hard scattering)
reduces the production of small $x$ partons by splitting, which in
turn are responsible for soft production, leading to a kinematic
depletion of soft particles if hard scattering occured in the event.
Similar to cross-section fluctuations, in~\cite{mcglinchey2016} the
notion of weakly interacting or ``shrinking'' nucleon is explored to
explain events with a high $p_T$ jet present, and predictions are made
for centrality-dependent jet yields in $p$+Au, $d$+Au and $^3$He+Au
collisions at RHIC energies.

Some heavy ion experiments in the meantime tried to modify their Glauber
calculations with bias factors~\cite{ppg160}, or published $R_{pA}$
with different $\omega$ parameters of the Glauber-Gribov
model~\cite{atlas2016b}.  The ALICE experiment chose a different path
by publishing nuclear modifications in terms of the purely
experimental ``event activity'' rather than turning it into event
geometry using a model that is not directly verifiable.

\section{Direct photons to the rescue?}

Let us summarize our findings.  The Glauber
model and the centrality determination based on it works well when two
large ions collide.  This is not surprising: even if a few nucleons
suffer ``extreme'' collisions, the regular soft particle production from
the average binary collisions dominates the event (see
also~\cite{reygers2007} and the correlations in~\cite{david2013}).  
Also, production of
high $p_T$ direct photons is well understood in $p$+$p$ at all
available energies.  Finally, independent of centrality, the high
$p_T$ photon
$R^{\gamma}_{AA}$ is consistent with unity in A+A, modulo some small
(and experimentally distinguishable) effects, listed earlier.

Now let us assume that the physics mechanisms in A+A are a superset of
the mechanisms in $p$+A -- there is no new physics in $p$+A that
wouldn't be present in A+A, albeit possibly suppressed by much larger
effects present only in A+A.  It then follows, that if photons prove to be
a ``standard candle'' in A+A, they will be standard candle in $p$+A,
too.  So far all measurements indicate that photons indeed {\it are}
standard candle in A+A, their yield is not modified from the expected
one in those cases, where centrality (and $N_{coll}$) is unambiguous.
If so, then there is little reason to assume that $R^{\gamma}_{pA}$ will be
modified (differ from unity) in $p$+A.

This provides an opportunity to test {\it a posteriori} any model or
procedure aimed to provide geometry ($N_{coll}$) related information in $p$+A
collisions.  The lithmus test is whether the photon nuclear
modification factor $R^{\gamma}_{pA}$
calculated with it is consistent with unity -- for all centralities and
in the entire high $p_T$ range --  or not.  Note that in light
of Fig.~\ref{fig:ppg160_ppbiasdata} this second condition is also very
important.  If $R^{\gamma}_{pA}$ deviates from unity significantly in
any direction, the model is very likely biased.  Clearly, our test 
doesn't provide any guidance how to contruct geometry/centrality models or
procedures.  However, it gives a decisive test whether they are viable
or not.


\end{document}